# Adhesion, Stiffness and Instability in Atomically Thin MoS$_2$ Bubbles.

David Lloyd[1], Xinghui Liu[2], Narasimha Boddeti[2], Lauren Cantley[1], Rong Long[2], Martin L. Dunn[3] and J. Scott Bunch[1,4*]

[1]Boston University, Department of Mechanical Engineering, Boston, MA 02215 USA

[2]University of Colorado, Department of Mechanical Engineering, Boulder, CO 80309 USA

[3]Singapore University of Technology and Design, Singapore, 487372

[4]Boston University, Division of Materials Science and Engineering, Brookline, MA 02446 USA

*e-mail: bunch@bu.edu

## Abstract

**We measured the work of separation of single and few-layer MoS$_2$ membranes from a SiO$_x$ substrate using a mechanical blister test, and found a value of 220 ± 35 mJ/m$^2$. Our measurements were also used to determine the 2D Young's modulus ($E_{2D}$) of a single MoS$_2$ layer to be 160 ± 40 N/m. We then studied the delamination mechanics of pressurized MoS$_2$ bubbles, demonstrating both stable and unstable transitions between the bubbles' laminated and delaminated states as the bubbles were inflated. When they were deflated, we observed edge pinning and a snap-in** **transition which are not accounted for by the previously reported models. We attribute this result to adhesion hysteresis and use our results to estimate the work of adhesion of our membranes to be 42 ± 20 mJ/m$^2$.**

Adhesive forces play an important role in shaping the mechanical behavior of atomically thin materials such as graphene or molybdenum disulfide, MoS$_2$. These forces keep the material clamped to the substrate, and also influence how the membrane folds[1], slides[2], and peels[3]. An understanding of adhesion in these materials is important in the fabrication of nanoelectromechanical systems[4], flexible electronic devices[5], graphene origami[1,6], graphene separation membranes[7], and stacked heterostructures formed from 2D materials. Atomically thin crystals may also provide a fruitful system in which to study novel features of friction and adhesion present only at the nanoscale[2,8–10]. In terms of device performance, adhesive forces determine the maximum strain 2D materials can support which is important in designing stretchable electronic devices[11] and pressure sensors[12].

The study of bubbles formed by atomically thin sheets has proven to be useful for discovering the adhesive and mechanical properties of these materials, and has allowed measurements of the adhesion energies[13], friction coefficient[14], and Young's modulus of graphene and other 2D materials[15]. In particular, Koenig et al. used a mechanical blister test to measure the adhesion energy between graphene and SiO$_x$ of ~450 mJ/m$^2$. Like graphene, atomically thin MoS$_2$ is a mechanically exceptional material[16], whilst also being piezoelectric[11,17] and a direct gap semiconductor with a highly strain sensitive band gap[18–21]. A good understanding of the mechanical stiffness and adhesion to the substrate is therefore of particular importance to this material which has applications involving the interplay between adhesive and tensile forces.

In this paper, we measure the work of separation (sometimes referred to as the adhesion energy) between MoS$_2$ and the substrate by employing the same geometry as used in our previous work[7,13,22], in which we suspend mechanically exfoliated or chemical vapor deposition (CVD) grown membranes over cylindrical microcavities etched into a silicon oxide (SiO$_x$) substrate (Fig. 1a and 1b). The devices are then placed in a pressure chamber filled with a gas of pressure $p_0$, which gradually leaks into the cavities through the SiO$_x$ substrate until the internal pressure $p_{int}$ reaches that of the chamber ($p_{int} = p_0$). We used either N$_2$, Ar, H$_2$ or He gas which allowed us to choose a convenient leak rate of the gas into the microcavities. When the devices are removed from the pressure chamber the internal pressure ($p_{int}$) is greater than the external pressure ($p_{ext} = 1$ atm), and this pressure difference ($\Delta p = p_{int} - p_{ext} > 0$) causes the membrane to bulge up (Fig. 1c and 1d). For each charging pressure $p_0$ we measure the deflection $\delta$ and radius $a$ of the bubble using an atomic force microscope (AFM) after which the devices are returned to the pressure chamber at a higher $p_0$ and the process is repeated. We fabricated devices of 1-3 layer thickness by mechanical exfoliation, and made monolayer devices from CVD grown MoS$_2$ using a PMMA transfer method (see supporting information for details). We transferred 6 different growths to produce CVD samples *N1-6*, with each containing many individual devices. The $SiO_x$ substrates were $O_2$ plasma cleaned prior to transfer.

As can be seen in Fig. 1d-f, increasing $p_0$ causes $\delta$ to increase with $a$ initially remaining pinned at the radius of the cylindrical microcavity, $a_0$. After a critical pressure is reached ($p_0 \sim$ 600 kPa), the force from the pressure difference across the membrane overcomes the adhesive forces keeping the membrane clamped to the substrate, and delamination occurs in the form of a snap-out transition of the radius from 4.4 µm to 6 µm. After the snap-out transition, both $a$ and $\delta$ continue to gradually increase as $p_0$ is increased.

We begin by using our values for $p_0$, $\delta$ and $a$ to determine the Young's modulus of $MoS_2$ with a formula developed in Hencky's model for clamped pressurized membranes[23], which relates the pressure difference across the membrane $\Delta p$ to the deflection $\delta$ and radius $a$ by the formula,

$$\Delta p = \frac{K(v)E_{2D}\delta^3}{a^4} \quad (1)$$

with a Poisson's ratio $v = 0.29$[16], numerical constant $K(v) = 3.54$ and a two dimensional Young's modulus $E_{2D}$ equal to the bulk Young's modulus multiplied by the thickness of the material. The pressure difference, $\Delta p$, is calculated from $p_0$ by assuming isothermal expansion of a fixed number of ideal gas molecules from the initial volume of the cavity ($V_0$) to its final volume ($V_0 + V_b$), such that $p_0 V_0 = p_{int}(V_0+V_b)$. From Hencky's model, the volume created beneath the bubble can be found from the device geometry using the expression $V_b = C(v)\pi a^2 \delta$, and a numerical constant $C(v) = 0.522$.

We measured the $E_{2D}$ of 3 CVD samples ($N1$-$3$), and of exfoliated monolayer and trilayer flakes containing 2 and 16 devices respectively. Fig. 2a shows a plot of $\Delta p$ against $K(v) \delta^3/a^4$ for each of our CVD monolayer and bilayer devices in sample $N2$, including linear fits which are used to determine $E_{2D}$ for each device. The $E_{2D}$ of each device in these samples is plotted in Fig. 2b. In Fig. 2c we plot the mean $E_{2D}$ for each sample divided by the number of layers $n$ in the membranes in order to compare estimates for the $E_{2D}$ of a single MoS$_2$ layer. Error bars represent the standard deviation.

For our exfoliated devices we find an average $E_{2D}$ per layer of $190 \pm 35$ N/m, and for our CVD grown MoS$_2$ monolayers we find an average $E_{2D}$ of $128 \pm 20$ N/m. There is a low variance of $E_{2D}$ within each CVD grown sample, however there is a significant difference between the average $E_{2D}$ for each CVD sample. The discrepancy between CVD and exfoliated samples and among different CVD samples may be due to differences in defect densities[24,25] which occur during CVD growth, as an increased sulfur vacancy density[26] is predicted to lower $E_{2D}$ in MoS$_2$[27]. The average of all our exfoliated and CVD grown samples is $160 \pm 40$ N/m, which falls within the same range of values as found in previous studies[16,28,29], which we plot in Fig. 2c for comparison.

We next determined the work of separation, $\Gamma_{sep}$, using our values for $p_0$, $\delta$ and $a$, and a free energy model described in detail by others[30,31]. Briefly, we can write the total free energy of the system $F$ as,

$$F = \frac{(p_{int}-p_{ext})V_b}{4} + \Gamma\pi(a^2 - a_o^2) - p_oV_o \ln\left[\frac{V_o+V_b}{V_o}\right] + p_{ext}V_b \tag{2}$$

where $V_0$ is the initial volume of the cavity, $V_b$ is the additional volume created as the bubble expands. $\Gamma$ is the adhesion energy, which is equal to $\Gamma_{sep}$ in the case of delamination. The first two terms represent the elastic strain energy and the work to separate the membrane from the substrate respectively, and the final two terms account for the isothermal expansion of the gas.

When a device is removed from the pressure chamber, the bubble volume expands until the free energy of the system $F$ reaches a local minimum. We minimize $F$ with respect to $a$ by setting $dF/da = 0$ and using the relationship $p_0V_0 = p_{int}(V_0 + V_b)$. This yields the expression for the work of separation:

$$\Gamma_{sep} = \frac{5C}{4}\left(\frac{p_oV_o}{V_o+V_b(\delta,a)} - p_{ext}\right)\delta \tag{3}$$

with the constant $C(v) = 0.522$ for $v = 0.29$[16]. Using this expression, we can determine $\Gamma_{sep}$ of each device using the charging pressure of the pressure chamber $p_0$, and $\delta$ and $a$ of the bubble measured using an AFM. We can also substitute the pressure terms in Eq. 3 with Hencky's result in Eq. 1 which yields,

$$\Gamma_{sep} = \frac{5}{4} C K E_{2D} \left(\frac{\delta}{a}\right)^4 \qquad (4)$$

which holds for all devices which have started to delaminate ($a > a_0$). This allows $\Gamma_{sep}$ to be determined from $\delta$ and $a$ without knowing $p_0$, which avoids the long waiting times required for devices to reach equilibrium in the pressure chamber. For our exfoliated devices we calculated $\Gamma_{sep}$ using Eq. 4 (using the mean value of $E_{2D} = 190$ N/m per layer we found earlier for exfoliated samples), and used Eq. 3 to calculate $\Gamma_{sep}$ for our CVD devices where $p_0$ was well known.

We find no significant difference in $\Gamma_{sep}$ between single and few layer samples, or CVD and exfoliated samples (Fig. 3). By averaging over all samples we find the mean work of separation to be $\Gamma_{sep} = 220 \pm 35$ mJ/m$^2$, which is close to the value of $170 \pm 30$ mJ/m$^2$ measured for many layer MoS$_2$[32] and is in the same range of values as found for graphene[13,33–36].

The devices shown in Fig. 1d-f exhibit unstable delamination, whereby $a$ discontinuously increases from the initial radius $a_0$ when $p_0 \gtrsim 600$ kPa. The etched depth of the microcavities in that case was $d = 1500$ nm. We also fabricated devices with cavity depths of $d = 650$ nm, and again performed measurements of $\delta$ and $a$ at increasing $p_0$ (Fig. S9) using the method described earlier. With this cavity depth, the devices show no snap-out transition, and rather stably delaminate with $a$ continuously increasing from $a_0$. The difference in behavior in these two cases has been observed and modeled by others[31,37], and Bodetti et al found that the transition from unstable to stable delamination occurs when the parameter $S = 2V_b/V_0$ satisfies the condition $S > 1$ just before the point of delamination[31]. Reducing the well depth decreases the volume of the cavity relative to the volume of the bubble which increases $S$. By making various device geometries and finding $S$ from AFM measurements we confirmed empirically that this transition occurs in the range $0.74 < S < 1.11$, and we obtained the same value for $\Gamma_{sep}$ for both stable and unstable delamination (see supplementary info for details).

After the devices with $d = 1500$ nm (on sample $N2$) had been delaminated to their largest radii, they were left out in ambient conditions to deflate over the course of ~48 hours. During this time AFM scans captured $\delta$ and $a$ as the number of gas molecules $N$ decreases from the initial value of $N_0$ ($= p_0V_0/k_bT$). AFM cross sections of a bubble are shown in Fig. 4a during the inflation (increasing $N_0$) and deflation (decreasing $N$) of the device. Initially as the device is inflated, $\delta$ increases and $a$ remains pinned at $a_0$. When $p_0 \gtrsim 600$ kPa the snap-out transition occurs and $a$ jumps to a larger value, after which both $a$ and $\delta$ increase together as $N_0$ increases. When devices are left to deflate, $\delta$ decreases from an initial value of $\delta_0$, however $a$ now does not change from its radius at the beginning of deflation, which we refer to as the 'pinned radius' $a_p$. After the deflection of the devices reaches a critical value $\delta = \delta_c$ the devices undergo a snap-in transition where the radius jumps from $a_p$ to $a_0$, and $\delta$ continues to decrease to zero. Values for $\delta$ and $a$ throughout this process are shown in Fig. 4b, which shows devices deflating at a number of different $a_p$. Videos of the snap-out and snap-in transitions can be seen in the supporting information.

We can interpret this using the result derived in Eq. 4, which requires that after delamination the ratio $\delta/a$ remains constant, with the magnitude of this ratio being proportional to $\Gamma_{sep}^{1/4}$. We plot the line corresponding to this formula in Fig. 4b (upper dashed line) with the values of $E_{2D}$ and $\Gamma_{sep}$ determined earlier, and find our data for increasing $N_0$ follows this trend very well.

This formula is independent of whether $N$ is increasing or decreasing, so when our devices are left to deflate we should expect $\delta$ and $a$ to return along the same path as during inflation described by Eq. 4. As can be seen in Fig. 4a however, there is a significant difference in the geometry of the bubbles during inflation and deflation, which suggests some element of our system is irreversible.

We attribute the difference between inflation and deflation we see in our data to the widely observed phenomenon of *adhesion hysteresis*[34,38,39], whereby the energy required to separate the membrane from the surface $\Gamma_{sep}$ is greater than the energy returned to the system as the membrane re-adheres $\Gamma_{adh}$, with $\Gamma_{adh} < \Gamma_{sep}$. After making this simple modification (see supporting information for more details), our model now predicts that the device should remain pinned at radius $a_p$ until a snap-in transition occurs at a critical deflection determined by

$$\Gamma_{adh} = \frac{5}{4}CKE_{2D}\left(\frac{\delta_c}{a_p}\right)^4 \tag{5}$$

We perform a linear fit of our measurements of $\delta_c$ and $a_p$ (lower dashed line in Fig. 4b) which yields an estimate of the work of adhesion for this sample to be $\Gamma_{adh} = 14 \pm 5$ mJ/m$^2$. Multiple measurements of $\Gamma_{adh}$ with the same device show that this measurement is repeatable over many cycles (Fig. S3b in the supporting information). We performed measurements on a total of 5 CVD grown samples (*N2-6*) and found the mean work of adhesion for all our samples to be $42 \pm 20$ mJ/m$^2$, with $\Gamma_{adh} < \Gamma_{sep}$ in every device. Fig. 4c shows a comparison between the works of separation and adhesion for 3 of these samples (*N2-4*). $\Gamma_{adh}$ varied noticeably between samples, with sample means falling in the range 14 - 63 mJ/m$^2$ (Fig. S5 in the supporting information).

Our measurements of $\Gamma_{adh}$ show that as little as one tenth of the energy required to separate the membrane from the substrate ($\Gamma_{sep}$ ~ *220* mJ/m$^2$) is recovered as the membrane at the edge of the bubble re-adheres to the substrate. We used Raman spectroscopy to measure the membrane strain distribution around our devices before and after snap-in (see supporting information for details), and found that whilst some energy was dissipated in the form of residual strain transferred to the membrane, this can only account for <10% of the dissipation that produces a difference between $\Gamma_{adh}$ and $\Gamma_{sep}$. This strain may also dissipate some energy through frictional sliding as the membrane changes its length on the surface of the substrate[14].

Adhesion hysteresis is a commonly observed phenomenon[40] which has previously been observed in nano-indentation measurements of graphene[34], and the fraction of the energy dissipated in our system is comparable with the hysteresis observed in elastomers[41]. The behavior of our devices is also analogous to the related phenomenon of contact angle hysteresis seen in liquid bubbles[39], and constant contact area pinning during unloading has been seen previously between two adhered solid spheres[42]. Surface roughness and chemical heterogeneity on the surface can produce contact angle and adhesion hysteresis[40,43], and a further contribution in our system could be the finite time over which deflation occurs. This could mean that the membrane does not have time during the measurement to re-conform fully to the surface or re-make the bonds which were made before the device delaminated[44,45]. This would result in the system being in a transient non-equilibrium state during the measurement, which is a common cause of thermodynamic irreversibility and adhesion hysteresis[40,46,47]. Our method of finding $\Gamma_{sep}$ also involves subjecting the membranes to high external pressures prior to measurement, which could improve their conformation to the substrate and thereby enhance $\Gamma_{sep}$ relative to $\Gamma_{adh}$.

We have measured the work of separation of single and few layer MoS$_2$ fabricated by CVD and mechanical exfoliation, and found a value of $\Gamma_{sep} = 220 \pm 35$ J/m$^2$. We also measured the Young's modulus, and found that $E_{2D} = 160 \pm 40$ N/m for a single MoS$_2$ layer. Bulge testing provides a complimentary method to nanoindentation to determine $E_{2D}$, and our results are in the same range of values as reported in previous studies. We demonstrated snap-out and snap-in instabilities, which mechanically amplify small changes in pressure and could be used for pressure sensing. Finally we observed bubble edge pinning, analogous to contact angle hysteresis observed in liquids, and used Raman spectroscopy to provide evidence that the trapping of strain energy after the snap-in transition can account for some but not all of the hysteresis. We measured a $\Gamma_{adh}$ which was significantly lower than $\Gamma_{sep}$, which may affect the performance of nanomechanical switches made from atomically thin materials[48,49]. The distinction between $\Gamma_{adh}$ and $\Gamma_{sep}$ we have observed here is an important consideration in the analysis of bubbles formed under atomically thin crystals[15,50,51], and in the design of folded 3D structures made from 2D sheets[1,6].

**Supporting information**

Supporting information includes details of CVD growth and characterization, the effect of membrane pre-tension, the full set of work of adhesion and separation data, the free energy model including the effect of adhesion hysteresis, contact angle measurements of a bubble during deflation, the trapping of strain around the edge of the devices, the effect of membrane slipping on our $E_{2D}$ calculations, data from devices which exhibit stable delamination, additional measurements of deflating devices, the full set of Young's modulus data, and videos of the snap transitions.

**Acknowledgments:**

This work was funded by the National Science Foundation (NSF), grant no. 1054406 (CMMI: CAREER, Atomic Scale Defect Engineering in Graphene Membranes), a grant to L. Cantley by the NSF Graduate Research Fellowship Program under grant no. DGE-1247312, and a BUnano Cross-Disciplinary Fellowship to D. Lloyd. We thank Chuanhua Duan for use of the high speed camera.

**Fig. 1**

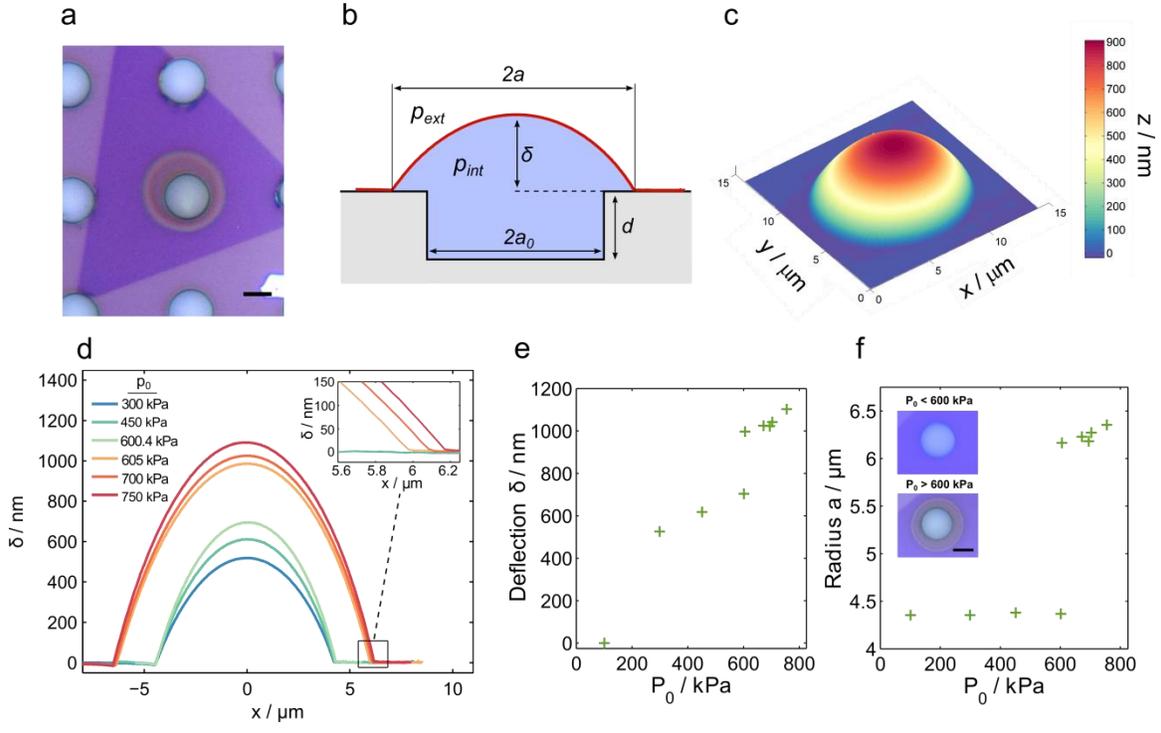

*Fig.1 a) Microscope image of a delaminated device (scale bar is 5μm). b) Device schematic. c) AFM image and d) AFM cross sections. e) Deflection δ and f) radius a plotted against input pressure $p_0$. Inset microscope images show a device before and after snap-out (scale bar is 5μm).*

**Fig. 2**

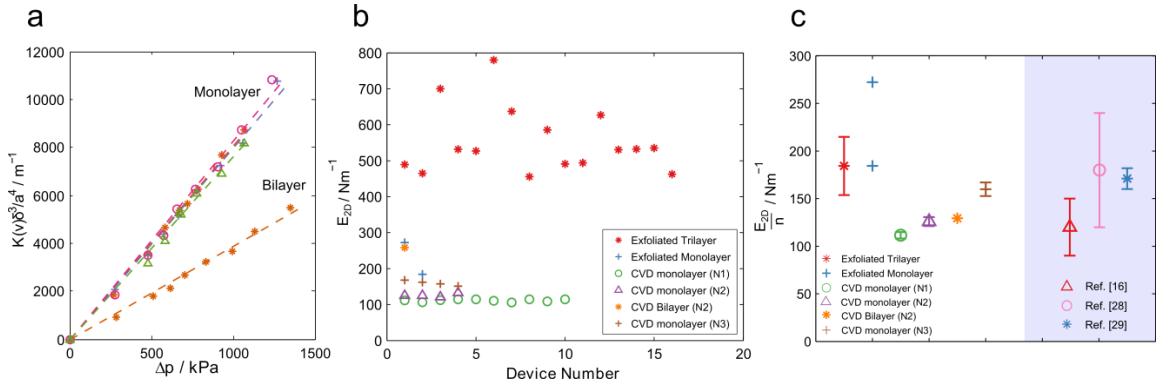

*Fig. 2 a) Plots for CVD monolayer and bilayer devices (different symbols/colors represent each device), with linear fits (dashed lines) used to find $E_{2D}$. b) $E_{2D}$ for each device in our exfoliated samples, and three of our CVD samples (N1-3) c) $E_{2D}$ divided by number of layers n for each sample. Data points and error bars represent the mean and standard deviation respectively for each sample. Results from nanoindentation measurements in references 16, 28 and 29 are plotted for comparison.*

**Fig. 3**

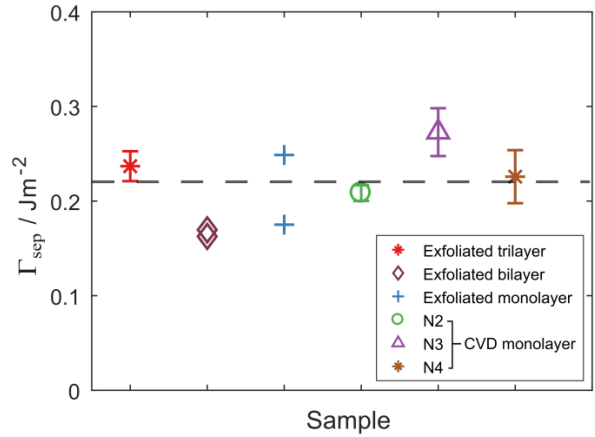

*Fig.3 Work of separation of membranes of 1 to 3 layer thickness. The data includes measurements of CVD monolayer devices from three separate growths and transfers (N2-4). Several devices are measured per sample, with data points and error bars representing the means and standard deviations respectively. For samples with fewer than 3 measurements the data points represent each device measured. The dashed line marks the mean of the 6 samples.*

**Fig. 4**

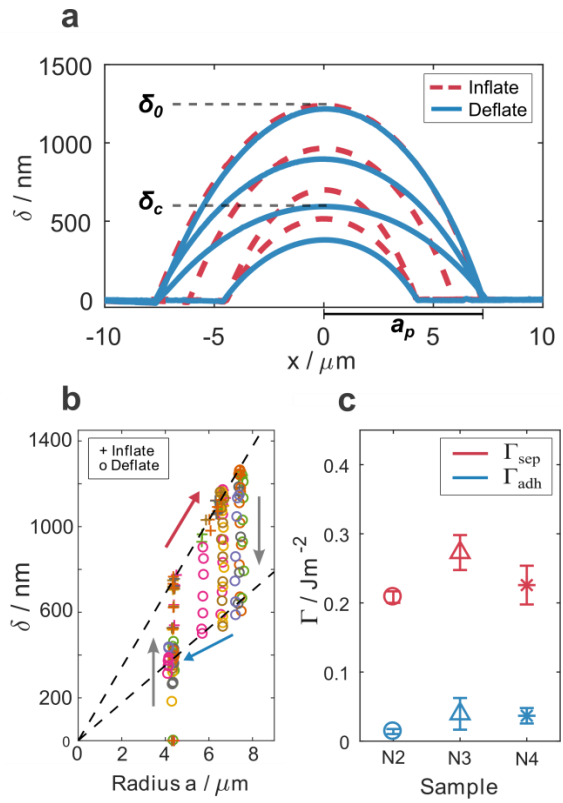

*Fig. 4 a) AFM cross sections of a device during inflation (increasing N) and deflation (decreasing N). Arrows mark the snap transitions. b) δ and a of devices during inflation and deflation. Different colors represent different devices on sample N2. More data can be found in the supplementary information which is not shown here for reasons of clarity. Red and blue arrows mark snap-out and snap-ins respectively. The upper and lower dashed lines correspond to solutions to Eq. 4 and Eq. 5 respectively. c) A comparison between the works of separation and adhesion for samples N2-4. Data points and error bars represent the means and standard deviations respectively of all the devices measured on each sample.*

# Supporting Information

# Adhesion, Stiffness and Instability in Atomically Thin MoS$_2$ Bubbles.

David Lloyd[1], Xinghui Liu[2], Narasimha Boddeti[2], Lauren Cantley[1], Rong Long[2], Martin L. Dunn[3] and J. Scott Bunch[1,4*]

[1]Boston University, Department of Mechanical Engineering, Boston, MA 02215 USA

[2]University of Colorado, Department of Mechanical Engineering, Boulder, CO 80309 USA

[3]Singapore University of Technology and Design, Singapore, 487372

[4]Boston University, Division of Materials Science and Engineering, Brookline, MA 02446 USA

*e-mail: bunch@bu.edu

# 1. Growth and characterization

Devices were grown by chemical vapor deposition (CVD) according to a method described in an earlier paper[1]. The devices were transferred over the etched microcavities using a PMMA dry transfer method. Immediately prior to transfer the $SiO_X$ wafers were $O_2$ plasma cleaned for 15 mins to remove any surface contamination. Before annealing off the PMMA layer at 340 °C, the devices were left in a vacuum desiccator for > 3 days to allow any gas trapped in the microcavities to leak out. Monolayers were identified by their optical contrast, and their Raman and photoluminescence (PL) spectra (Fig. S1). The separation between the $E^1_{2g}$ and $A_{1g}$ Raman modes was 20.3 cm$^{-1}$, and the A exciton peak in the PL spectrum was located at 1.88 eV, which demonstrates that the membrane was single layered[2,3]. The $E^1_{2g}$ peak position is later used to determine the residual membrane strain.

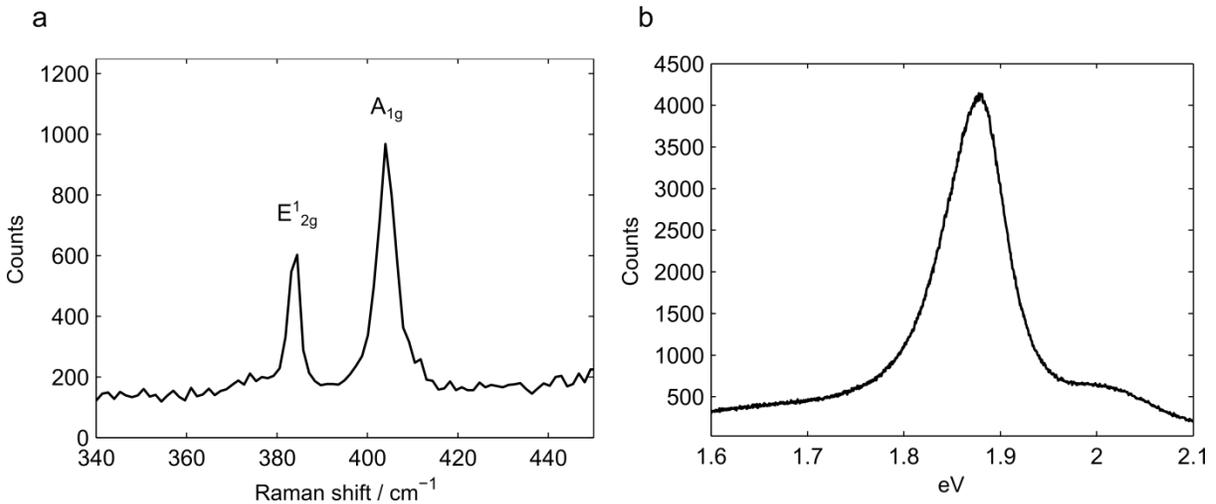

Fig. S1 a) The Raman and b) PL spectrum of a suspended single layer $MoS_2$ device with zero pressure difference across the membrane.

# 2. The effect of membrane pre-tension

Even when there is no pressure difference across the membrane there is usually a residual pre-strain observed in suspended devices, due either to the transfer procedure or the membrane sticking to the sidewalls of the cavity[4]. We can estimate the pre-tension in our membranes by using photoluminescence spectroscopy. In an earlier paper[1] we showed that the band-gap in monolayer $MoS_2$ reduces when biaxial strain is applied, at a rate of - 99 meV/%. We took a PL spectrum of a device with no pressure difference across the membrane (Fig. S1), meaning any observed strain would correspond to the pre-strain. We can then convert this to a pre-tension using the formula[5],

$$\sigma_0 = \frac{E_{2D}\varepsilon_0}{1-\nu} \qquad (S1)$$

Our devices have a pre-strain of $\varepsilon_0 < 0.002$ which corresponds to a pre-tension of $\sigma_0 < 0.2$ N/m, which is comparable to previously reported values for atomically thin membranes in this geometry[4,6]. Campbell 1956 [5] showed that when the non-dimensional parameter,

$$P = \frac{\Delta p a E_{2D}^{1/2}}{\sigma_0^{3/2}} \qquad (S2)$$

satisfies the condition $P > 100$, Hencky's formula in Eq. 1 is correct to within 5%. Most of our data points were taken in a high enough pressure range to satisfy this condition. For instance for the data presented in Fig. 2a, $P = 100$ when $\Delta p = 350$ kPa. Since nearly all of our data was taken with $\Delta p > 350$ kPa we use Eq. 1 to calculate $E_{2D}$, and neglect the effect of the pre-tension.

## 3. Work of separation

The full set of data used to produce means and standard deviations of each sample in Fig. 3 of the main text is shown in Fig. S2. Each data point represents the measured value of $\Gamma_{sep}$ for an individual device of a given sample.

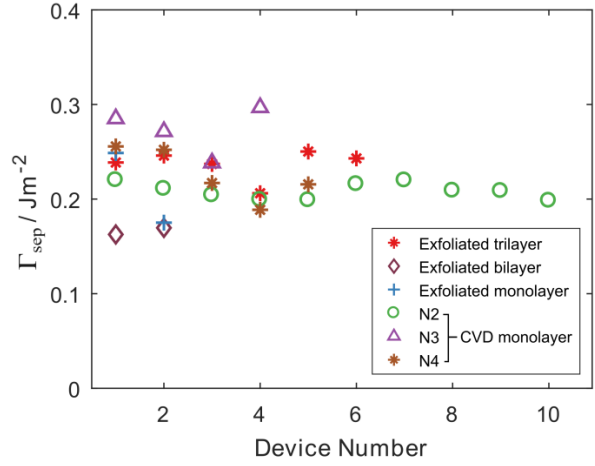

Fig. S2. All $\Gamma_{sep}$ data used to calculate means and standard deviations of each sample in Fig. 3.

## 4. Free energy model including adhesion hysteresis

We can interpret the results described in Fig. 4 of the main text using the free energy model described in Eq. 1. Taking the derivative of $F$ with respect to $a$, and substituting the pressure terms for the Hencky's result in Eq. 3 yields,

$$\frac{dF}{da} = 2\pi a \left[ \Gamma - \frac{5}{4} C K E_{2D} \left(\frac{\delta}{a}\right)^4 \right] \tag{S3}$$

Setting this formula equal to zero to find the radius at which the free energy is minimized leads to,

$$\Gamma_{sep} = \frac{5}{4} C K E_{2D} \left(\frac{\delta}{a}\right)^4 \tag{S4}$$

The constants $C$ and $K$ depend only on the Poisson's ratio $v$, and their values for various 2D materials are tabulated in Table S1.

|  | Poisson's Ratio $v$ | $K(v)$ | $C(v)$ |
|---|---|---|---|
| MoS$_2$ | 0.29 | 3.54 | 0.522 |
| Graphene | 0.16 | 3.09 | 0.524 |
| hBN | 0.22 | 3.28 | 0.523 |
| Black Phosphorus | 0.4 | 4.07 | 0.519 |

Table S1. Values for constants *C(v)* and *K(v)* for several 2D crystals, calculated using Hencky's solution.

We plot the relationship described by Eq. S4 in Fig. S4a with a value of $\Gamma_{sep} \sim 220$ mJ/m$^2$ and find our data fits this relationship very well. This formula is independent of whether *N* is increasing or decreasing, so when our devices are left to deflate we should expect $\delta$ and *a* to return along the same path as during inflation, and described by Eq. S4.

We can explain the difference between inflation and deflation we see in our data as a result of *adhesion hysteresis*, whereby the energy required to separate the membrane from the surface $\Gamma_{sep}$ is greater than the energy returned to the system as the membrane re-adheres $\Gamma_{adh}$, with $\Gamma_{adh} < \Gamma_{sep}$.

For changes of the device radius *Δa*, we now have:

$$\Gamma = \begin{cases} \Gamma_{adh}, & \Delta a < 0 \\ \Gamma_{sep}, & \Delta a > 0 \end{cases} \tag{S5}$$

As the device inflates and *Δa >0,* the free energy of the system is minimized according to Eqs. 4 and 5, with $\delta_0/a_p \sim \Gamma_{sep}^{1/4}$. When deflating the radius of the device will only decrease when *dF/da >0* for *Δa <0* (with $\Gamma=\Gamma_{adh}$), in order for the free energy to be minimized. From examining Eq. S3 and considering that $\Gamma_{adh} < \Gamma_{sep}$, this will only occur when $\delta$ has decreased from $\delta_0$ to below the critical value of $\delta = \delta_c$ after which the device radius can reduce in the form of a snap-in transition. Since the radius cannot decrease until $\delta_c$ is reached, the bubble edge remains pinned at $a_p$. The critical deflection $\delta_c$ marks the point where *dF/da = 0* for *Δa <0* (i.e. $\Gamma=\Gamma_{adh}$), and from using Eq. S3 we can see that this occurs when the relationship,

$$\Gamma_{adh} = \frac{5}{4} C K E_{2D} \left(\frac{\delta_c}{a_p}\right)^4 \tag{S6}$$

is satisfied. This corroborates with what we see in Fig S3a, in which the value of $\delta_c$ is roughly proportional to $a_p$ for the devices measured. We can estimate the value of $\Gamma_{adh}$ by fitting this relationship to the values of $\delta_c$ and $a_p$ of devices just before the snap in transition occurs, and we plot this line of best fit in Fig. S3a which corresponds to $\Gamma_{adh} \sim$ 14 mJ/m². We checked the repeatability of our measurements of $\Gamma_{adh}$ by repeating the experiment 6 times on a single device, which resulted in a mean and standard deviation of 13 mJ/m² and 5 mJ/m² respectively (Fig. S3b).

These arguments are best seen graphically in terms of the free energy landscape plotted as a function of radius in Fig S4b, c and d. In the absence of adhesion hysteresis, as the pressure inside the device decreases and the devices deflate, the free energy minima moves to a smaller radius (Fig. S4b). The path taken by our devices is shown in Fig. S4c, and clearly shows the devices not following the local minima in the free energy. By introducing adhesion hysteresis into the model (Fig. S4d), $\Delta F$ is calculated using $\Gamma_{sep}$ for $\Delta a > 0$ and $\Gamma_{adh}$ for $\Delta a < 0$, which results in the device radius remaining trapped in a local minima as the device deflates. The radius only changes when $dF/da > 0$ for $\Delta a < 0$ which only happens when Eq. S6 (Eq. 5 in the main text) is satisfied.

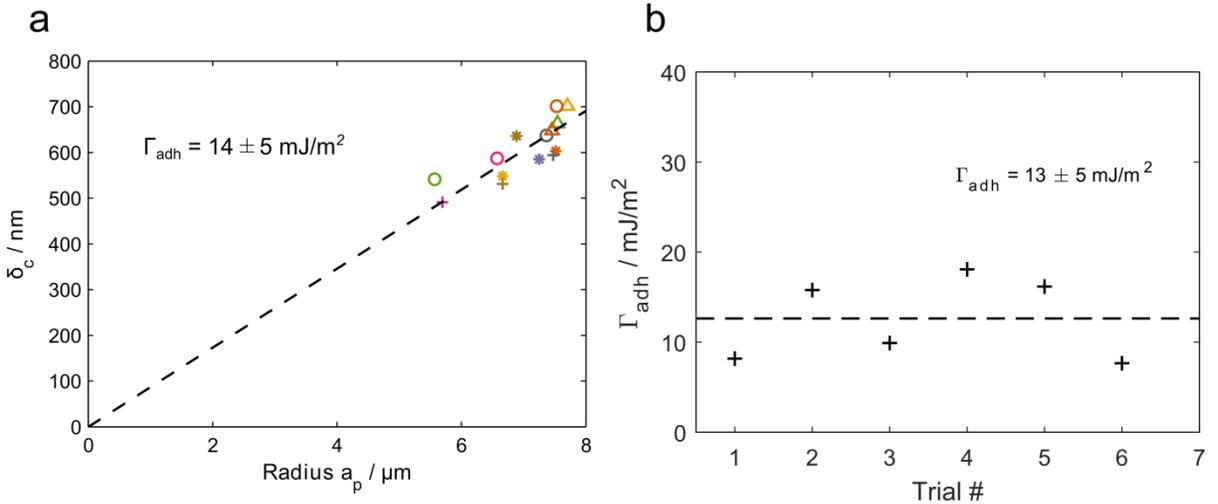

Fig. S3 a) Data for devices measured on sample *N2*, showing the values of $\delta_c$ and $a_p$ just before snap-in used to calculate $\Gamma_{adh}$. Each color/symbol represents a different device. b) Multiple measurements of a single device at a number of different pressures showing repeatability. Dashed line represents the mean adhesion value.

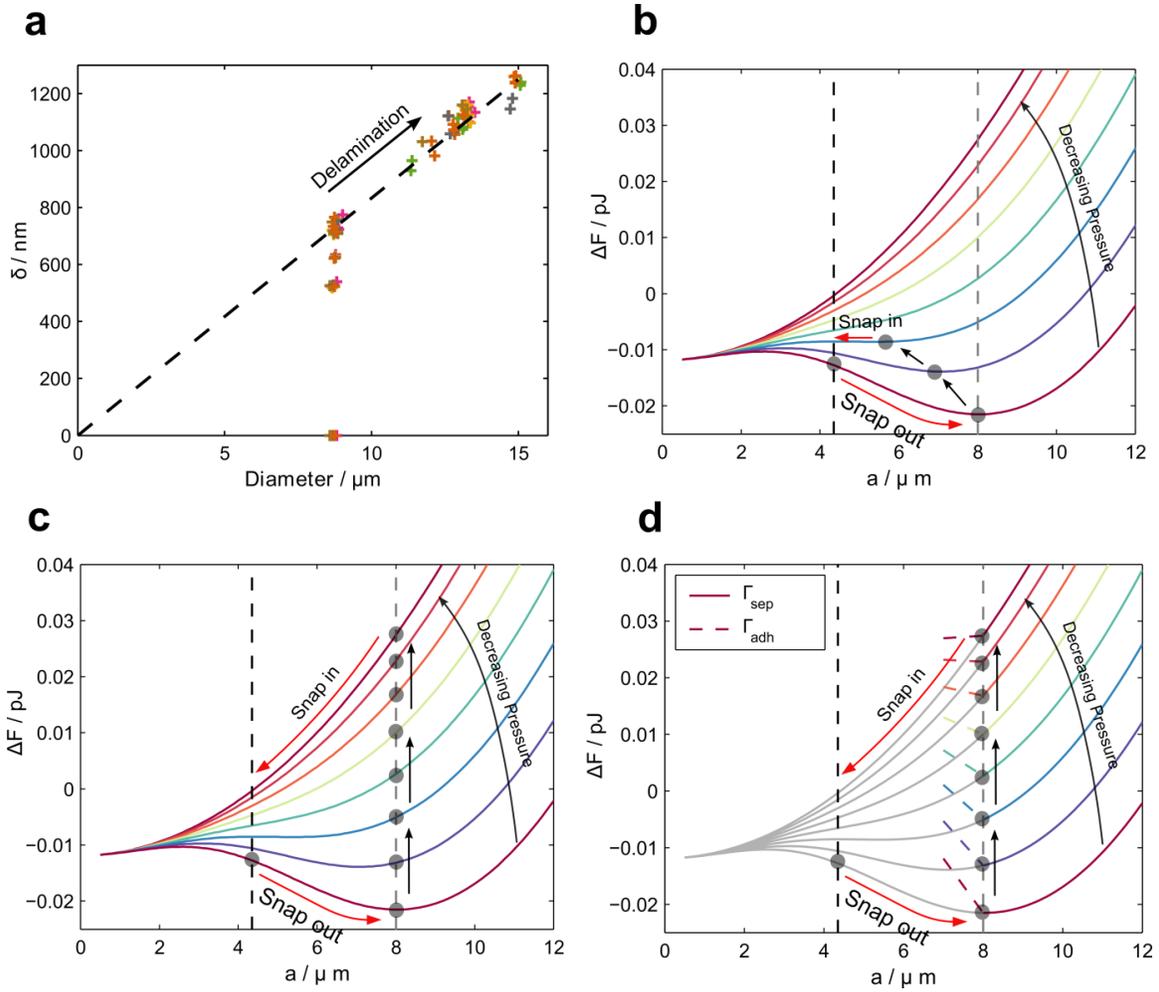

Fig. S4 a) As devices delaminate the ratio $\delta/a$ remains constant according to Eq. S4. b) The free energy landscape if there is no adhesion hysteresis. The device radius is that which minimizes the free energy, and the grey dots mark the path we would expect the device to take. c) The actual path our devices take, which appears to not minimize the free energy. d) The modified free energy landscape if $\Gamma_{adh} < \Gamma_{sep}$. As the device reduces its radius its free energy is determined by the dashed lines. The device is now trapped in a free energy minima and snap-in only occurs when the gradient of the dashed line is greater than zero.

To see if the work of adhesion varied between samples fabricated with the same method of CVD growth and transfer, we performed measurements of 5 different CVD samples (*N2-6*) with at least 4 devices measured per sample. Monolayer devices were delaminated and left to deflate, and AFM measurements of $\delta_c$ and $a_p$ taken just before snap-in were used to calculate the work of adhesion using Eq. S6. The data is presented in Fig. S5a, with each data point representing a measurement of $\Gamma_{adh}$ in a single device of a given sample. The mean and standard deviations of each sample are shown in Fig. S5b. Between different samples there is considerable variation in the mean work of adhesion, which suggests that factors such as the cleanliness of substrate or membrane which can vary from sample to sample may play significant roles in adhesion hysteresis. A few of the devices measured did not snap in completely from radius $a_p$ to $a_0$, but rather initially snapped in to an intermediate radius followed by a second snap in to $a_0$ (Fig. S9b). All the transitions between these states were unstable and occurred in less than one second.

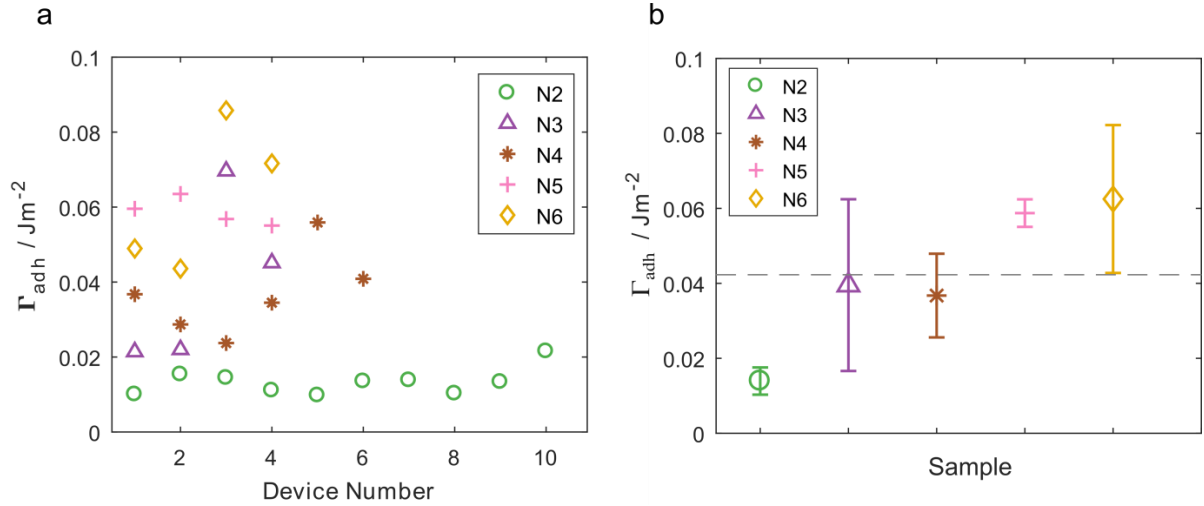

Fig. S5 a) Work of adhesion for every device measured in each sample. b) Mean and standard deviations of the work of adhesion in each sample. The dashed line represents the mean of the 5 samples.

## 5. Contact angle of bubbles during deflation

Instead of analyzing the snap-out and snap-in data in terms of $\delta$ and $a$, an analogous method is to measure the contact angle $\theta_c$ between the membrane and the substrate (see Fig. S6 inset) using an AFM. In Fig. S6 we plotted the contact angle against the radius of a device as it is inflated (black) and then left to deflate (red). As the device is inflated the contact angle increases until a critical value, at which point the device delaminates with the contact angle remaining constant. When the device is left to deflate the contact angle decreases at constant radius until another critical contact angle is reached, at which point the device undergoes the snap-in transition.

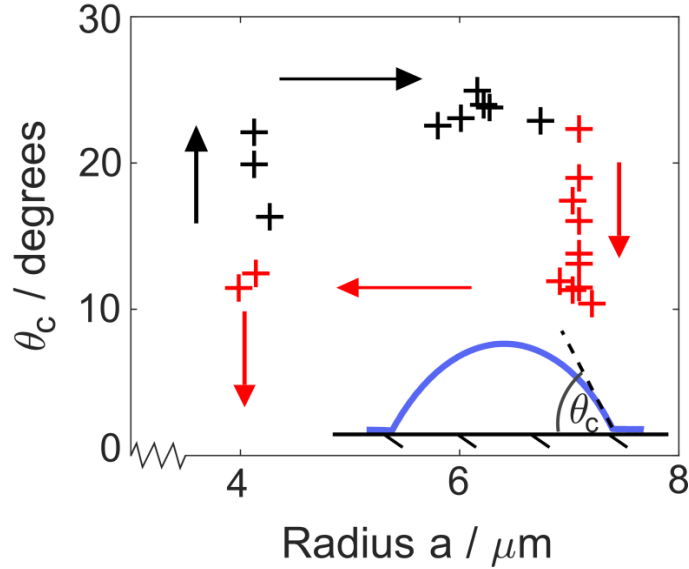

Fig. S6. The contact angle of a device during inflation (black) and deflation (red).

## 6. Strain trapping around the edge of the membrane

To investigate a possible mechanism for the observed adhesion hysteresis we used Raman spectroscopy to measure the strain distribution around our devices. The peak positions of the Raman modes in monolayer $MoS_2$ are known to be sensitive to strain[1,7], so by measuring how these peaks shift at different locations around the device we can build up an image of how strain is distributed. For these measurements we used the $E^1_{2g}$ peak to estimate the strain (Fig. S1), since it has a peak position which is strain sensitive and independent of doping effects.

Fig. S7a shows an AFM image of a device delaminated to $a_p \sim 7.5$ μm, which was then left to deflate and undergo the snap-in transition. A Raman map was then taken after snap-in (Fig S7b), with the strain calculated from the position of the $E^1_{2g}$ peak using the reported shift rate of $\sim 5$ cm$^{-1}$ / %[1,7]. A region of $\varepsilon \sim 0.5\%$ can be clearly seen around the circumference of where the delaminated bubble was before snap-in. This strain likely originates from the pressure induced radial strain at the edge of the bubble, which for these devices is ~1.5% (Fig. S8d). Using this upper bound of $\varepsilon \sim 1.5\%$ and the formula for the isotropic strain membrane energy density[8], $U = \frac{1}{2} E_{2D} \varepsilon^2$, we can estimate the energy stored in the strained regions to be $U \sim 20$ mJ / m$^2$, which can account for some but not all the energy dissipation which produces a difference between $\Gamma_{adh}$ and $\Gamma_{sep}$. The presence of strain in the membrane also implies some contribution of energy dissipation through friction as the membrane changes its length on the surface of the substrate[9].

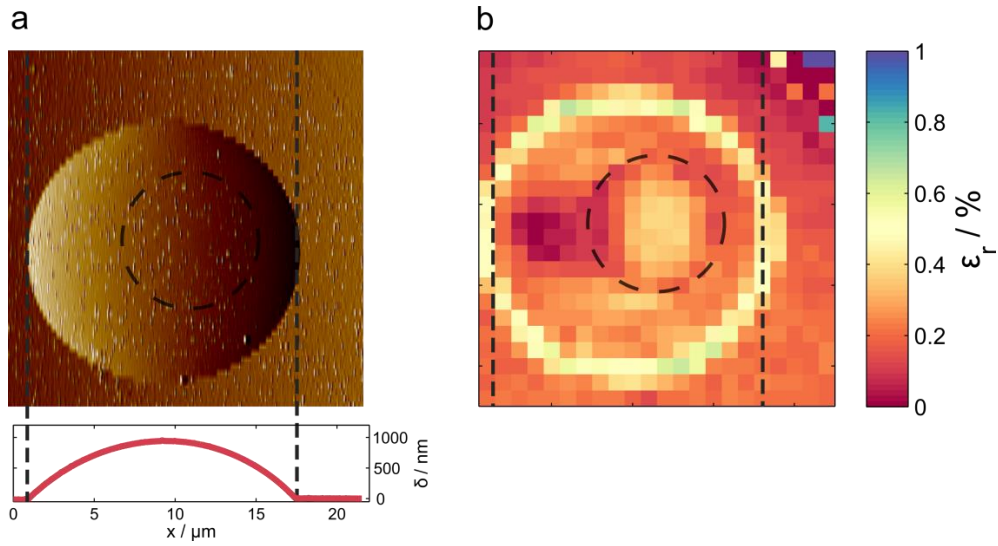

Fig. S7 a) AFM image (amplitude channel) of a delaminated device before the snap-in transition. The position of the microcavity is marked by a dashed circle. Below is a cross section of the device. b) Strain map of the same device after the snap-in transition when the device has fully deflated. Strain is calculated using the peak shift in the $E^1_{2g}$ Raman mode at each point. Each pixel is 1 x 1 μm and corresponds to a single Raman scan.

In order to observe the process by which this strain becomes 'trapped' in the membrane around the device, we took Raman line scans over a cross section of a device as it deflated and plotted the $E^1_{2g}$ peak position as a function of distance (Fig. S8a and S8b). Before each Raman scan we found the corresponding geometry of the device by taking an AFM image (Fig. S8c). Across the delaminated bubble region (marked by dashed lines) the peak shift abruptly increases at the edge of the bubble, followed by a gradual increase towards the center of the device. In Fig. S8d we used Hencky's solution to find the predicted strain profile across the device for its initial geometry (Fig. S8c red line) before deflation. In the model, the strain jumps from zero to purely radial tensile strain at the edge of the device, with the tangential component gradually increasing from zero to be equal to the radial component at the center. The $E^1_{2g}$ peak position depends on contributions of both the radial and tangential strain, so this model explains the profile seen in Fig. S8a.

Fig. S8b shows that a region of strain extends ~1.5 µm outside the edge of the bubble in the initial Raman scan (red line). As the device deflates and the radius remains pinned the peak shift across the delaminated region of the membrane reduces as it becomes less strained, however the region of strain outside bubble remains roughly constant throughout deflation. These results show that the ring of strain in Fig. S7b is formed when the device initially delaminates, and that this strain does not relax as the device deflates and eventually snaps in.

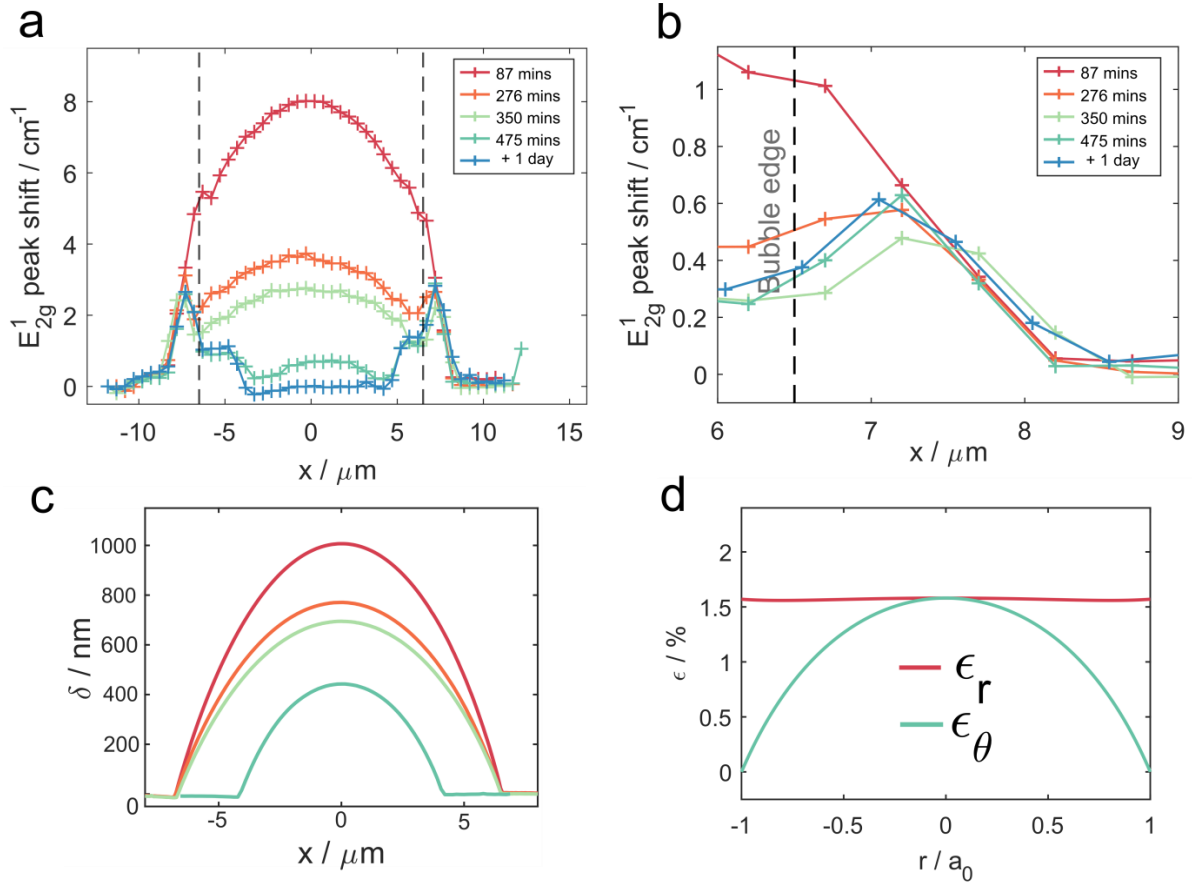

Fig. S8 a) Raman line scans over a device over time as it deflates. Dashed vertical lines mark the edge of the delaminated bubble. b) A zoomed in version of a) focusing on the edge region of the device. c) AFM cross sections of the device at each time, using the same color scheme as in a). d) Radial ($\varepsilon_r$) and tangential ($\varepsilon_\theta$) components of the strain as a function of radius for this device's initial geometry before deflating, calculated using Hencky's model with values for δ and a taken from the red curve in c).

## 7. The effect of the slipping of the membrane on $E_{2D}$ calculations

The strain at the edge of the bubble introduces extra slack into the membrane of bubble, which may affect our measurements of $E_{2D}$. We can estimate the effect this has on our measurements by integrating the strain over the strained region at the edge of the bubble in Fig. S8b to find the total extra slack, $\Delta L$, added to the bubble membrane. We can write the slack added to the membrane as,

$$\Delta L = \int_0^{x_1} \varepsilon(x)dx \tag{S7}$$

The initial measurement in Fig. S8b (red line color and labeled '87 mins') shows that the peak shift linearly decreases from ~ 5.5 cm$^{-1}$ around the edge of the device to ~0 cm$^{-1}$ at 1.5 μm outside the device radius, so we take $x_1$ = 1.5 μm. To find $\varepsilon(x)$ we take $\varepsilon$ ~ 1.5 % at the edge of the device (Fig. S8d) and use the linear strain profile seen in Fig. S8b, which leads to $\varepsilon(x)$ ~( 0.015/1.5) $x$ μm$^{-1}$. This gives $\Delta L$ ~ 11 nm over a device radius of 6.5 μm. This reduces the pre-strain by ~ 0.0017 which is about the same as the initial pre-strain. We therefore take this change to be negligible in to the pressure range we are studying due to the arguments made in section 2.

## 8. Stable delamination devices

Devices of well depth $d$ ~ 650 nm were fabricated that exhibited stable delamination (Fig. S9). These devices showed the same hysteric behavior as our other devices. To calculate the work of separation of these devices we used Eq. 4 in the main text with AFM measurements of δ and a, and used the mean $E_{2D}$ of all our CVD devices of 128 N/m. We measured 3 devices over 4 different pressures, and found a value of $\Gamma_{sep}$ = 207 ± 19 mJ/m$^2$. We also measured the work of adhesion of the device shown in Fig. S9b, which we found to be $\Gamma_{adh}$ = 40 mJ/m$^2$.

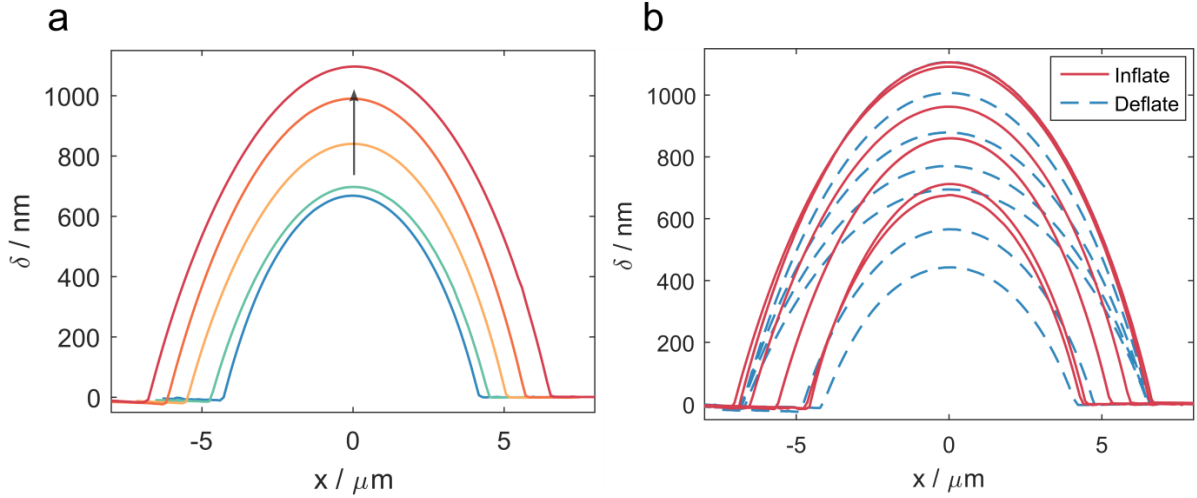

Fig. S9 a) Stable delamination with increasing pressure. b) A device which delaminates stably with increasing pressure, but shows adhesion hysteresis upon deflation. This device snapped in to an intermediate step before fully re-laminating to the substrate.

## 9. Additional snap-in data

Fig. S10 shows the complete data set for our snap-in measurements presented in Fig. 4b of the main text. This data was taken using an AFM in tapping mode. To confirm that the forces from to the AFM tip were not affecting our results, we measured the snap-in of a device as it deflated by using solely optical measurements. We took sequential PL spectra at the center of the device as it deflated, where the membrane is under biaxial strain. In an earlier paper[1] we found that the PL peak red-shifts under biaxial strain by -99 meV/%, so PL measurements allow us to measure the biaxial strain $\varepsilon$ in the device. We can also measure the radius $a$ of the device as it deflates using an optical microscope. Using these values for $a$ and $\varepsilon$ we can estimate the deflection of the device using the formula,

$$\varepsilon = \sigma(\nu)\left(\frac{\delta}{a}\right)^2 \tag{S8}$$

where $\sigma(\nu)$ is a numerical constant which depends only on Poisson's ratio $\nu$, and in this case $\sigma = 0.709$. We measured a deflating device using the non-contact optical method, after which we re-inflated the device to the same pressure and used the AFM to measure the geometry of the device as it deflated. We compare the results of these two methods in Fig. S10b, and find very similar results in the two cases. The device appears to snap-in at a slightly lower $\delta$ in the AFM measurements, however this is likely due to the long scan times (~3 min) required to take a PL spectrum meaning that we couldn't measure the device right at the moment before snap-in.

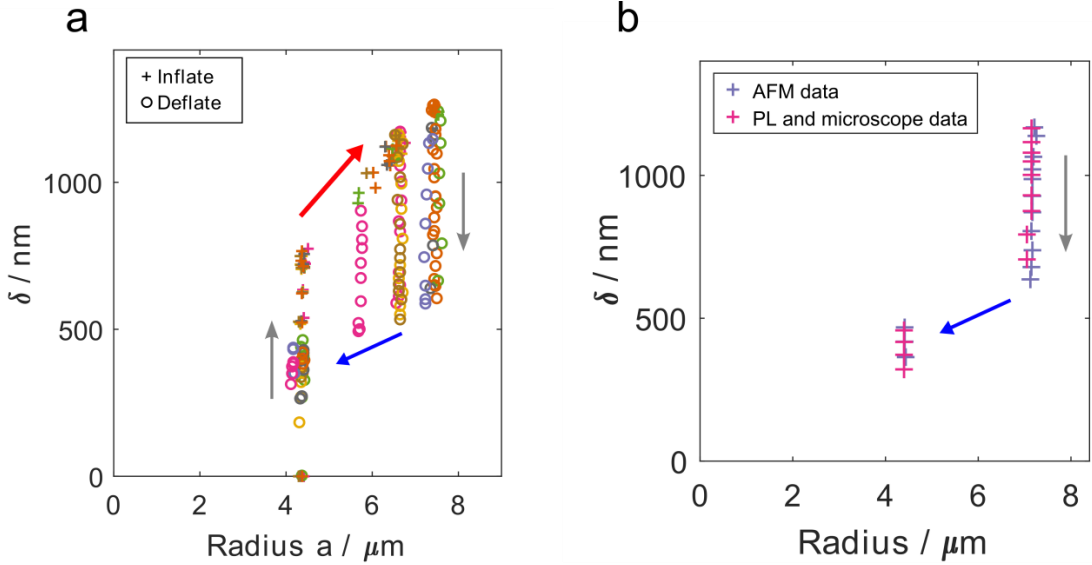

Fig. S10 a) Complete data containing all data points of results presented in Fig. 4b in the main text. Each color represents a different device. b) Comparison of snap-in transitions measured optically or by AFM. For optical measurements $a$ is determined using an optical microscope with a 100$x$ objective, and $\delta$ is determined from the PL peak position and Eq. S8.

## 10. Young's modulus

Fig. S11 shows the complete data set used to calculate the Young's modulus for each device in Fig. 2b in the main text.

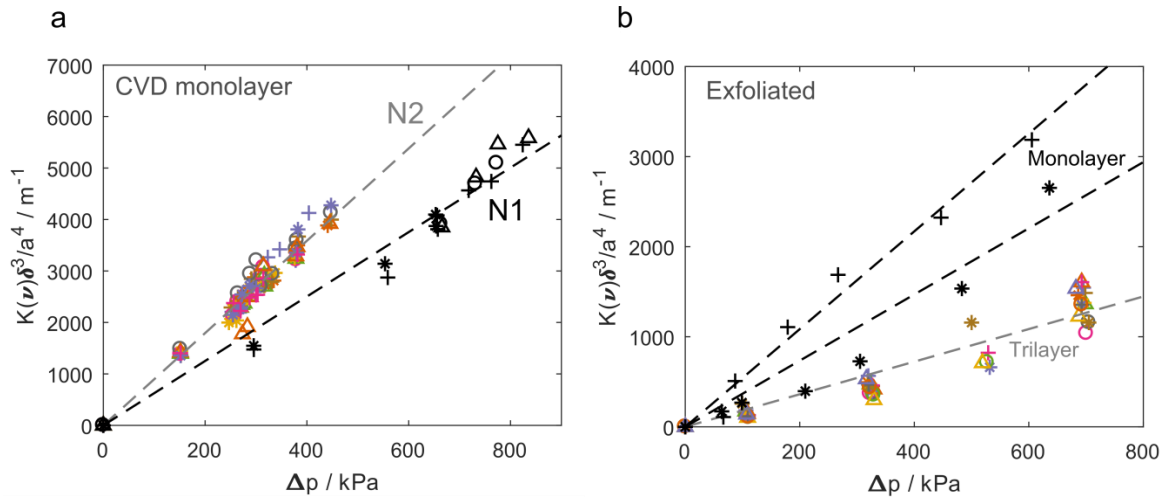

Fig. S11 a) CVD monolayer devices from sample *N1* and *N2*. b) Exfoliated monolayer and trilayers devices. Dashed lines are plotted for each of the sample means reported in Fig. 2c of the main text. Different color/symbols represent different devices.

## 11. Videos of snap transitions

Video 1 shows the snap-in transition of a deflating device taken with a high speed camera. The snap-in transition occurs faster than the frame rate of the camera (0.5 ms). Video 2 shows a device in a pressure chamber with a quartz window, allowing us to observe a delaminated device as the chamber pressure is increased and decreased (video speed is 4x). For the first half of the video the external pressure is increased, with the delaminated device snapping-in at ~6 s. During the second half of the video the pressure is decreased, with the device snapping-out at ~30 s.